\documentclass[preprint,aps,prl]{revtex4}

\usepackage{graphicx}
\begin{document}
\title{Size of Orbital Ordering Domain Controlled by the Itinerancy of the 3$d$ Electrons in a Manganite Thin Film}
\author{Y.~Wakabayashi$^{1,2}$, H.~Sagayama$^3$, T.~Arima$^3$, M.~Nakamura$^4$\footnote{Present address: RIKEN, Hirosawa, Wako 351-0198, Japan}, Y.~Ogimoto$^{5,6}$, Y.~Kubo$^5$, K.~Miyano$^{5,6}$ and H.~Sawa$^{1,7}$}
\address{$^1$Photon Factory, Institute of Materials Structure Science, High Energy Accelerator Research Organization, Tsukuba 305-0801, Japan\\
$^2$Graduate School of Engineering Science, Osaka University, Toyonaka 560-8531, Japan\\
$^3$IMRAM, Tohoku University, Sendai 980-8577, Japan\\
$^4$Department of Applied Physics, University of Tokyo, Tokyo 113-8586, Japan\\
$^5$Research Center for Advanced Science and Technology, University of Tokyo, Tokyo 153-8904, Japan\\
$^6$CREST, Japan Science and Technology Agency, Honcho 4-1-8, Kawaguchi 332-0012, Japan\\
$^7$Department of Applied Physics, Nagoya University, Nagoya 464-8603, Japan
}

\date{\today}
\begin{abstract}
An electronic effect on a macroscopic domain structure is found in a strongly correlated half-doped manganite film Nd$_{0.5}$Sr$_{0.5}$MnO$_3$ grown on a (011) surface of SrTiO$_3$. 
The sample has a high-temperature (HT) phase free from distortion above 180~K and two low-temperature (LT) phases with a large shear-mode strain and a concomitant twin structure. One LT phase has a large itinerancy (A-type), and the other has a small itinerancy (CE-type), while the lattice distortions they cause are almost equal.
Our x ray diffraction measurement shows that the domain size of the LT phase made by the HT-CE transition is much smaller than that by the HT-A transition, indicating that the difference in domain size is caused by the electronic states of the LT phases.
\end{abstract}

\maketitle

Macroscopic domain structures, such as magnetic ones, martensites, and lamella or gyroid structures of liquid crystals, abound in various materials. While the properties of each domain's constituents are often dominated by short-range interactions such as exchange interactions, the macroscopic shape and size of the domain are usually dominated by long-range interactions, such as static electromagnetic interactions or stress, because the effect of the long-range interactions accumulates as the domain grows. 

Many manganites also show lamella-type domain structures in their orbital ordered phases\cite{Li01JAP,Tokunaga08PRB}. Although the orbital ordering itself is a manifestation of the strong correlation among the electrons\cite{Tokura00Science}, the domain structure is primarily dominated by local stress\cite{Li01JAP}.
We have found, in contrast to this, a significant change in the macroscopic domain structure caused by a change in the itinerancy of the $3d$ electrons in a Nd$_{0.5}$Sr$_{0.5}$MnO$_3$ thin film on a SrTiO$_3$ (011) substrate (NSMO/STO) through x ray diffraction measurements under magnetic fields. Since this change in itinerancy is made by applying the magnetic field, most of the extrinsic effects such as chemical deffects are excluded. From materials science standpoint, a good understanding of the domain structure gives us further insights into the macroscopic material properties of strongly correlated systems, because the percolative phenomena are often observed in various strongly correlated materials\cite{Dagotto05NJP}. 

\begin{figure}
\includegraphics[width=7cm]{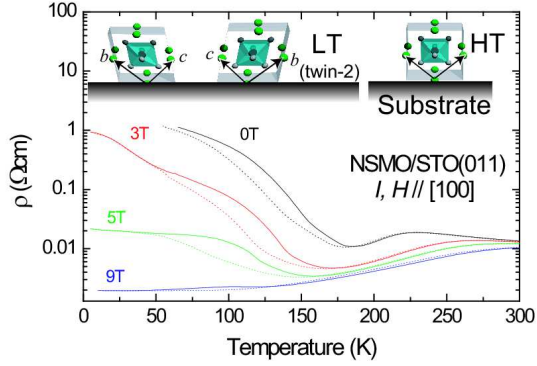}
\caption{(Color online) Temperature dependence of the resistivity in various magnetic fields. Solid curves and dashed curves show the results on warming and cooling runs, respectively. Inset: Schematic view of HT phase and two twin-2 structures with shear mode strain.}
\label{fig:properties}
\end{figure}

The orbital states of $e_g$ electrons are best studied by examining structural properties because of the strong coupling between the $e_g$ electrons and the lattice\cite{SRXS}. This coupling makes a martensitic strain when charge and orbital ordering occur\cite{Podzorov01PRB}, and the strain may cause a phase coexistence in some cases\cite{Ahn04Nature}. Our previous study on NSMO/STO zero field\cite{Ogimoto05PRB,Wakabayashi06PRL,Wakabayashi08JPSJ} shows that this film has a ferromagnetic (FM) phase below 210~K, $x^2-y^2$-type ferroorbital ordered phase (A-OO phase) below $T_{\mbox{\scriptsize  A}} =$ 180~K, and $3x^2-r^2/3y^2-r^2$ antiferroorbital ordered phase (CE-OO phase) below $T_{\mbox{\scriptsize  CE}} =$ 160~K. There are two kinds of twin structures. One is related to the film growth --- that is, to the tilt of the $a$*-axis in [011] direction by $\pm$0.4$^\circ$ --- (twin-1), and the other happens at $T_{\mbox{\scriptsize  A}}$ at which the $b$ and $c$ lattice parameters split (twin-2, see the inset of Fig.~\ref{fig:properties}). Throughout this paper, we use cubic notation of the lattice parameters for the sake of direct comparison with references [\onlinecite{Wakabayashi06PRL,Wakabayashi08JPSJ}]. From the viewpoint of structural deformation, we call the phase having the relation $b=c$ the HT phase (paramagnetic and FM phases), and $b\neq c$ the LT phase (A-OO and CE-OO phases). The HT-LT transition at $T_A$ is a kind of martensitic transition\cite{Podzorov01PRB}, which changes the lattice parameters without making a diffusion of the atoms, as well as a Jahn-Teller transition. The metal-insulator transition temperature $T_{MI}$ coincides with $T_{\mbox{\scriptsize  A}}$ in a heating run in zero field.

In the present study, we observed clear phase coexistence of the HT phase and LT phase under the conditions of high magnetic field and low temperature. This remnant HT phase is found to originate from the large stress induced at the twin-2 domain boundary, and the LT-phase domain size is estimated from the detailed measurement of the LT-HT ratio as a function of temperature and magnetic field, as shown later.

Epitaxial films were grown by the pulsed laser deposition (PLD) method\cite{Ogimoto05PRB}. The thickness of the NSMO film for this study was 80~nm, while no essential thickness dependence ranging from 50 nm to 110 nm was observed in transport and magnetic properties.\cite{Ogimoto05PRB}. 
First, we investigated transport properties of the sample with a standard four probe method with 5T-Magnetic Property Measurement System and 9T-Physical Property Measurement System (Quantum Design). Figure~\ref{fig:properties} displays temperature dependence of resistivity measured along [100] axis by applying various magnetic fields with field cooling and field warming processes. The film has a clear first-order insulator-metal phase transition. $T_{MI}$ and the resistivity in the insulating phase decreases as the field increases, which is naively understood in terms of the stabilization of the ferromagnetic spin arrangement and the resulting increase of the hopping probability.

X ray diffraction experiment with photon energies of 12~keV and 14~keV in magnetic fields was carried out at BL-3A and BL-16A1 of the Photon Factory, KEK, Japan. A large two-circle diffractometer attached to an 8~T superconducting magnet was installed to these beamlines for this experiment; in the present configuration, the magnetic field was applied perpendicular to both [011] and the scattering vector, which depends on the domain because the NSMO/STO has intrinsic twin structures, twin-1 and twin-2. 

\begin{figure}
\includegraphics[width=7cm]{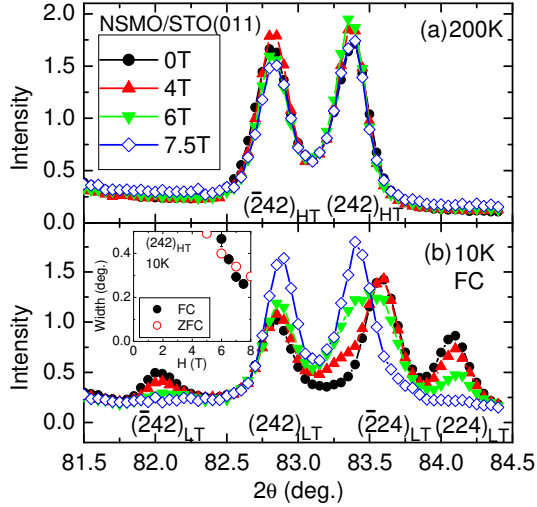}
\caption{(Color online) Intensity profile around 242 Bragg reflection measured at (a) 200~K, and (b) 10~K. The intensity was normalized so that the integrated intensity is constant when the magnetic field is applied. The inset for panel (b) shows the magnetic field dependence of the peak width of (242)$_{HT}$ component measured at 10K with the FC and ZFC procedures.}
\label{fig:profile}
\end{figure}

Figure~\ref{fig:profile} shows the intensity profiles around 242 Bragg reflection measured at (a) 200~K and (b) 10~K with various magnetic fields. A magnetic field was applied in the FM phase, and field cool (FC) was made for the measurements. Each plot shows the integrated intensity with respect to the diffractometer angle $\omega$, and the figure shows the integrated intensities as a function of $2\theta$. 
 The magnetic field significantly affects the peak profile at 10~K, while little change with the magnetic field is observed at 200~K. In panel (a), only two peaks corresponding to twin-1 are seen. According to the lattice parameters at zero field\cite{Wakabayashi06PRL}, the lower angle peak can be assigned to ($\bar{2}42$)$_{\mbox{\scriptsize  HT}}$ and ($\bar{2}24$)$_{\mbox{\scriptsize  HT}}$  and  the higher angle to ($242$)$_{\mbox{\scriptsize  HT}}$ and ($224$)$_{\mbox{\scriptsize  HT}}$. Here, the suffix HT represents the indices in the lattice parameters for the HT phase. The relation $b=c$ in the HT phase makes the peak positions of $hkl$ and $hlk$ identical. In panel (b), there are four peaks related to both twin-1 and twin-2 in the zero-field profile. They are assigned to ($\bar{2}42$)$_{\mbox{\scriptsize  LT}}$, ($242$)$_{\mbox{\scriptsize  LT}}$, ($\bar{2}24$)$_{\mbox{\scriptsize  LT}}$ and ($224$)$_{\mbox{\scriptsize  LT}}$, from lower to higher angles, respectively. These low-temperature profiles had little temperature dependence up to 60~K. The magnetic field prominently affected these peaks. At 4~T, a bump appears at the left side of the ($\bar{2}24$)$_{\mbox{\scriptsize  LT}}$ peak, and the intensity of $(224)_{\mbox{\scriptsize  LT}}$ reflection decreases. The bump increases its intensity as the magnetic field increases, and at 7.5~T, the profile becomes almost identical to that at 200~K. The peak width, or the inverse correlation length, of the magnetic field-induced HT phase peak (242)$_{\mbox{\scriptsize  HT}}$ is shown in the inset of Fig.~\ref{fig:profile} (b) as a function of the magnetic field. The width becomes broader, or the correlation length becomes shorter, as the field decreases to 4~T.

\begin{figure}
\includegraphics[width=7cm]{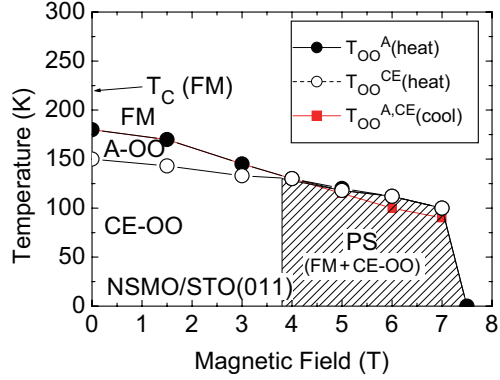}
\caption{(Color online) Temperature-magnetic field phase diagram for NSMO/STO(011). The magnetic field was parallel to the [11\=1] or [1\=11] direction (depending on the domain). The magnetic field was applied in the FM phase, and transition temperatures were measured in both cooling and warming runs. PS represents the phase segregated state of FM and CE phases.}
\label{fig:Phase_Dia}
\end{figure}

Superlattice reflections corresponding to the CE-OO phase are observed on the reciprocal lattice points of $\sqrt 2 \times 2\sqrt 2 \times 2$ unit cell for the LT phase lattice parameters, as reported previously under zero field\cite{Wakabayashi06PRL}. These reflections are also observed in the magnetic field up to 7~T. The temperature dependence of the integrated intensity of (1/4 9/4 2)$_{\mbox{\scriptsize  LT}}$ superlattice and the $(224)_{\mbox{\scriptsize  LT}}$ Bragg reflections was measured in order to investigate the $T_{\mbox{\scriptsize  CE}}$ and $T_{\mbox{\scriptsize  A}}$ values, respectively, in various magnetic fields. Below 4~T, the transition temperature $T_{\mbox{\scriptsize  A}}$ differs from $T_{\mbox{\scriptsize  CE}}$, while they merge into one above this field. 
These features are summarized into a phase diagram shown in Fig.~\ref{fig:Phase_Dia}.
The A-OO region decreases as the magnetic field increases. Above the crossing field, phase separation (represented by the shade) appears as seen in Fig.~\ref{fig:profile} (b). At 4~T, $(242)_{\mbox{\scriptsize  HT}}$ appears as a bump, $(242)_{\mbox{\scriptsize  HT}}$ and $(\bar{2}24)_{\mbox{\scriptsize  LT}}$ are comparable at 6~T, and the HT phase is dominant at 7.5~T.

\begin{figure}
\includegraphics[width=7cm]{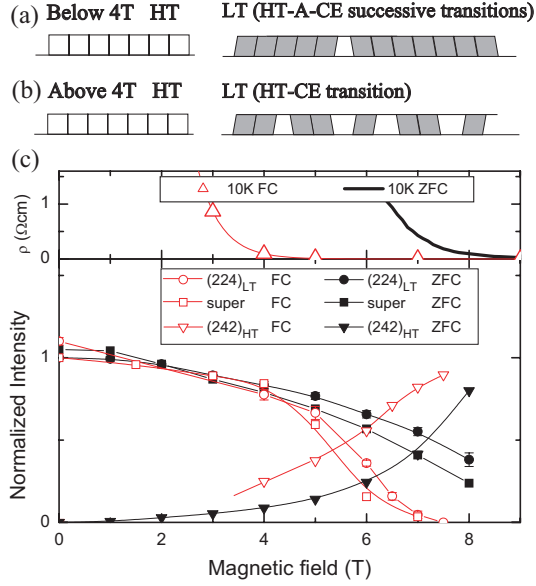}
\caption{(Color online) (a) Schematic view of the film at (left) high-temperature  and (right) low-temperature for the cooling field below 4~T, (b) those for the cooling field above 4~T, and (c)volume fraction of the LT phase estimated by (224)$_{\mbox{\scriptsize  LT}}$ intensity (circles) and the superlattice intensity (squares) measured at 10~K with field-cool (open symbols) and zero-field-cool (closed symbols) as a function of magnetic field, together with the HT phase fraction by (242)$_{\mbox{\scriptsize  HT}}$ intensity. Electric resistivity at 10~K in the field-cool (open triangles) and zero-field-cool (solid lines) is also presented.}
\label{fig:LT_Frac}
\end{figure}

Here we discuss how the phase separation state occurs in the film.
Either ferro- or antiferro-orbital ordering in a film system on (011) substrates requires the shear mode lattice distortion\cite{Wakabayashi08JPSJ} shown in the inset of Fig.~\ref{fig:properties}. 
This mode of distortion, called the $q_2$ mode, makes the lattice parameters $b$ and $c$ different. In an NSMO/STO film, twin-2 occurs at the HT-LT phase transition, as shown in Fig.~\ref{fig:profile}(b). In such a twinned system, there must be martensitic domain walls, which entail a large stress. In many bulk martensite transition systems, the stress is often released by changing the shape of the crystal or making dislocations\cite{martensite_text}. However, the former cannot happen in a film system because the substrate prohibits accumulation of the change in lattice parameters into a change in the size or shape of the whole sample. The latter also is not the case, because the maximum translational displacement caused by the HT-LT transition, calculated as the thickness of the film multiplied by the tilt angle, is about 4~\AA\/, less than the $d$-spacing of [$0\bar 11$]. Only zero or one dislocation plane is expected.
Therefore, a large stress that disallows the $q_2$ mode lattice distortion is induced around the domain wall. In turn, the orbital ordering is also suppressed. This situation is depicted in Figs.~\ref{fig:LT_Frac} (a) and (b).

According to this scenario, no induced HT phase is expected in twin-2 free films. One such example is Pr$_{0.55}$(Ca$_{0.8}$Sr$_{0.2}$)$_{0.45}$MnO$_3$ film on (011) surface of [(LaAlO$_3$)$_{0.3}$(SrAl$_{0.5}$Ta$_{0.5}$O$_3$)$_{0.7}$] substrate, which has a charge-ordering transition at 160~K \cite{Takubo05PRL}. Our recent structural study on this film shows neither twin-2 nor any signature of an induced HT phase\cite{Wakabayashi08EPJ}. This result supports the formation of the interfacial HT phase.

One of the intriguing facts seen in the phase diagram is the coincidence of the magnetic field where the phase coexistence starts and the crossing field of $T_{\mbox{\scriptsize  A}}$ and $T_{\mbox{\scriptsize  CE}}$. This coincidence suggests that the 3$d$ electronic state affects the martensitic transition. 
In the rest of this paper, we discuss the relationship between $3d$ electronic state and the phase coexistence.

Under the assumption that the HT phase at the low temperatures is located in the domain wall, its volume fraction is determined by two factors, the thickness and the number of the domain walls, the latter is related to the size of LT domains corresponding to twin-2, while the former is the size of the HT phase itself.
The twin-2 domains are formed at the HT-LT phase transition point. Due to the large elastic energy associated with the domain wall, it is not likely that the domains once formed can merge at lower temperatures, which implies that the number of domains is fixed at the transition temperature. The almost zero HT phase under zero field at low temperature suggests that the domain wall is narrow. As the magnetic field increases, the ferromagnetic HT phase is energetically more favored hence the wall thickness grows. The gradual increase of the HT volume fraction under the ZFC condition, presented in Fig.~\ref{fig:LT_Frac}(c), and the growth of the correlation length of HT phase for the ZFC case, in Fig.~\ref{fig:profile} inset, both support the picture.

Figure~\ref{fig:LT_Frac}(c) shows the magnetic field dependence of both HT and LT phase fractions in the FC and ZFC processes. The sudden increase (decrease) of the HT (LT) phase and the associated coexistent behavior at 4~T observed in the FC process is puzzling at first sight, because the lattice distortions in A-OO phase and CE-OO phase are alike and thus energetically similar, which leads to similar thickness of the interfacial HT phase for above and below the cooling field of 4~T. We thus have to seek another mechanism, the increase of the number of domains, as the cause. 
 Based on the phase diagram, the LT domain size made at $T_{\mbox{\scriptsize  A}}$ is found to be much larger than that made at $T_{\mbox{\scriptsize  CE}}$.  This situation is illustrated in Figs.~\ref{fig:LT_Frac} (a) and (b).

Various differences are observed in resistivity and structural property between the ZFC and the FC processes. As can be seen in Fig.~\ref{fig:LT_Frac}(c), the magnetic field dependence of the resistivity measured at 10~K with the FC process is much smaller than that with the ZFC process under high magnetic fields, whereas they are the same at 0~T.
The magnetic filed dependences of the volume ratio between the HT phase and LT phase measured with the FC and ZFC processes significantly differ to each other (Fig.~\ref{fig:LT_Frac} (c)), while those of the width of the (242)$_{\mbox{\scriptsize  HT}}$ peak measured at 10~K are very similar (inset of Fig.~\ref{fig:profile} (b)). All these features are well understood with the above mentioned mechanism. 
The peak width of the induced HT phase reflects the thickness of the boundary. It is dominated by the relationship between the elastic energy and the energy gain from the HT-LT phase change. Thus, the same thickness is expected for the same temperature and magnetic field regardless which procedure is used to reach the condition. This explains the similar peak widths.
 The resistivity difference implies that the FC procedure gives larger fraction of the metallic phase, i.e., the HT phase than the ZFC process in high magnetic fields. This result is consistent with the volume fraction obtained by the Bragg peak intensities. As described above, this difference is interpreted to be caused by the different density of the domain wall.

Finally, we discuss the origin of the difference in domain size. We propose a model where the domain size of the orbital ordered phase is dominated by the range of the electronic movement or itinerancy. The A-OO phase has high electric conductivity in the $c$-plane\cite{Nakamura04JPSJ}, which means that the $e_g$ electrons in the A-OO phase travel around nearly freely in the plane. At the HT-A transition temperature, one large orbitally ordered plane is established in a short time, and it distorts neighboring layers and forms a large domain. In contrast, the electrons in the CE-OO phase form zig-zag chains and the electron movement is confined in one-dimensional chains. At the HT-CE transition temperature, the zig-zag chain along the [110]$_{\mbox{\scriptsize  HT}}$ and [101]$_{\mbox{\scriptsize  HT}}$ directions grows independently and forms large numbers of small domains. Note that both transitions are first-order transitions; therefore, once the nucleation starts, the transition runs over a large volume with the speed of lattice vibration. 
In summary, the difference in ordering process is regarded as difference in speed of crystallization of electronic system that alters the domain size.

The authors thank Dr. T.~Fujii and Prof. A.~Asamitsu for the use of 9 T-PPMS system, and Dr. K.~Hagita for valuable discussion. This work was supported in part by grant-in-aid for scientific research from MEXT, Japan, and by the Sumitomo foundation.

\end{document}